\documentclass[prb,twocolumn,showpacs,preprintnumbers,superscriptaddress]{revtex4-1}
\usepackage[usenames,dvipsnames]{color}
\usepackage{amssymb} 
\usepackage{times}
\usepackage{graphicx}
\usepackage{graphicx}
\usepackage{dcolumn}
\usepackage{datetime}
\usepackage{soul}
\usepackage{subfigure}
\usepackage[bookmarks, colorlinks=true, plainpages = false,citecolor = red, urlcolor = black, filecolor = blue]{hyperref}
\usepackage{multirow} 
\usepackage[english]{babel}
\usepackage{graphicx}
\usepackage{amsmath}
 
\begin{document}
\title{Switching of a Quantum Dot Spin Valve by Single Molecule Magnets}
 
\author{Fatemeh Rostamzadeh Renani} 
\affiliation{Department of Physics, Simon Fraser University, Burnaby, British Columbia, Canada V5A 1S6}

\author{George Kirczenow} 
\affiliation{Department of Physics, Simon Fraser
University, Burnaby, British Columbia, Canada V5A 1S6}

\date{\today}

\begin{abstract}\noindent
We explore theoretically the spin transport in nanostructures consisting of a
gold quantum dot bridging nonmagnetic electrodes and two Mn$_{12}$-Ph single
molecule magnets (SMMs) that are thiol-bonded to the dot but are not in direct
contact with the electrodes. We find that reversal of the magnetic moment of
either SMM by the application of a magnetic field leads to a large change  in
the resistance of the dot, i.e., a strong spin valve effect. We show that this
phenomenon arises from a novel physical principle: The spin-dependent molecular
orbitals that extend over the dot and both SMMs change drastically when the
magnetic moment of either SMM is reversed, resulting in a large change in the
conduction via those orbitals. The same physics may also be responsible for the
spin valve phenomena discovered recently in carbon nanotube quantum dots with
rare earth SMMs by Urdampilleta, Klyatskaya, Cleuziou, Ruben and Wernsdorfer
[{\em Nature Mater.} {\bf 10,}  502506 (2011)].

\end{abstract}
 \pacs{85.75.Bb,75.50.Xx, 85.75.-d, 85.35.Be}

\maketitle

Molecular spintronics \cite{Sanvito2006} combines two active fields of study: molecular electronics and spintronics.
The ultimate goal of this field is to control the electronic spin and charge on the molecular scale where quantum effects emerge.
In the search for new molecular spintronic nano devices, single molecule magnets (SMMs) \cite{Gatteschi_book2006} appear to be natural candidates.\cite{SpinValveProposal} 
A SMM is a molecular-size nano-magnet that exhibits quantum behaviour such as quantum tunneling of the magnetization (QTM) \cite{Gatteschi_book2006}, Berry phase interference \cite{berry_phase}, the Kondo effect \cite{kondo2009}, and 
magnetoresistance phenomena. \cite{SpinValveProposal, SupramolecularSPinValve2011} 
Spin valve-like effects have also been predicted in SMMs with magnetic electrodes.\cite{conventional_SV1, conventional_SV2}In a recent advance, experimental detection of spin valve-like behavior has been reported\cite{SupramolecularSPinValve2011} in a system with non-magnetic electrodes and SMMs coupled to a (non-magnetic) carbon nanotube quantum dot. It was suggested\cite{SupramolecularSPinValve2011} that this behavior was due to the magnetic moments of individual SMMs bound to the dot reversing at different  values of the applied magnetic field as the field was swept. It was proposed\cite{SupramolecularSPinValve2011} that this would result in changes in the resistance of the dot due to modulation of the spin transport through the dot, i.e., a spin valve effect. However, to our knowledge, no relevant quantitative theory has as yet been reported. Thus the physics responsible for the observed novel behavior\cite{SupramolecularSPinValve2011} has not been definitively identified.
 
Here we explore spin valve phenomena in devices with a pair of SMMs bound to a
non-magnetic quantum dot bridging non-magnetic electrodes theoretically for the first time. 
Our results reveal
that spin valve functionality in such devices can arise from current carrying
electronic resonant states that extend over the whole device, including the
quantum dot and both SMMs that are bound to it. When the magnetic moment of one
of the SMMs is reversed, these states, being spin-dependent, are strongly
modified. Consequently the resistance of the device changes. Thus the system
exhibits a spin valve effect. Thus our findings demonstrate a novel principle
for spin valve operation that differs fundamentally from that based on
exchange-modulated tunnel barriers that was put forward in Ref.
\onlinecite{SupramolecularSPinValve2011}.  
We also show that this spin valve
phenomenon is not limited to devices with carbon quantum dots and rare
earth-based SMMs as in the recent experiment.\cite{SupramolecularSPinValve2011}
We predict  such spin valve behaviour also in other devices having different properties. These include, for example, the much better electronic transport characteristics (absence of Coulomb
blockade \cite{PRB, PRBrapid}) exhibited by spin valves with gold quantum dots
and some transition metal-based single molecule magnets.

We consider SMMs based on Mn$_{12}$\cite{Lis}. 
Mn$_{12}$ contains four Mn$^{4+}$ ions surrounded by a nonplanar ring of eight Mn$^{3+}$ ions. The large spin\cite{S10}  S=10 
and magnetic anisotropy barrier (up to $\sim 6.1$ meV along the easy axis)\cite{expt_MAB} lead to high blocking temperatures ($\sim 3.5$ K) \cite{MnLigandEffectZagaynova} and long relaxation times.\cite{expt_RelaxationTime} 
These properties make the Mn$_{12}$ family promising candidates for molecular spintronics. 
However, Mn$_{12}$ deposition on different surfaces has met with varying degree of success.\cite{book_deposit_SMM_Surface2006, mannini2009, Kahle2012}
Nevertheless, 
Mn$_{12}$ has been chosen to explore the spintronics of the first SMM transistors experimentally and signatures of magnetic states were observed.
\cite{Heersche2006, moonHo2006}

\begin{figure}[t!]
\centering
\includegraphics[width=1\linewidth]{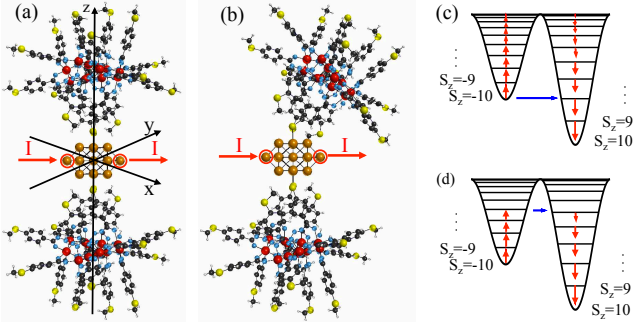}
\caption{(Color online) Geometry of the spintronic nano-device (a-b) and dynamics of the magnetization in single molecule magnets (c-d). The SMM's easy axes are (a) aligned and parallel to z-axis, (b) misaligned by 36$^{\circ}$. The red circles indicate the gold atoms coupled to leads.
Atoms are color labeled: gold (orange), manganese (red), carbon (grey), sulfur (yellow), oxygen (blue), and hydrogen (white). (c) In the thermally activated process (red arrows) the system overcomes the potential barrier by climbing up and down the complete ladders of spin energy levels. The blue arrow shows the quantum tunneling of the metastable ground state to the opposite side of potential barrier.
(d) Thermally assisted tunneling in which the quantum tunneling happens between energy levels near the top of the potential barrier that are populated through thermally activated processes. } 
\label{geometry} 
\end{figure}

We present quantum transport calculations that demonstrate a spin valve effect in nano-devices that consist of two Mn$_{12}$O$_{12}$(CO$_{2}$C$_{6}$H$_{4}$SH$_{3}$)$_{16}$ (henceforth Mn$_{12}$-Ph) SMMs coupled to a small gold quantum dot containing 15 gold atoms as shown in Fig. \ref{geometry} (a) and (b). Although the SMMs do not lie directly in the current path in these devices, we shall show that the relative orientation of their magnetic moments can strongly influence the electric current passing through the device.
Thus, we predict these systems to exhibit a strong spin valve effect.
We also found similar results for other gold nano-cluster sizes and geometries. 
Therefore, the spin-valve effect that we predict here does not depend critically on the gold nano-cluster size.
The non-magnetic leads are modeled as a small number of semi-infinite one-dimensional ideal channels that represent macroscopic electron reservoirs (as in previous studies of electron and spin transport through single molecules with gold contacts \cite{PRB, PRBrapid, idealLeads1, idealLeads2, idealLeads3}) to simulate moderately weak coupling between the nano-device and electrodes.

The geometries of our nano-devices are shown in Fig. \ref{geometry}(a) and (b). The Mn-containing magnetic cores of the Mn$_{12}$-Ph molecules are surrounded by organic ligands that are terminated by thiol-methyl (SCH$_{3}$) groups, the methyl (CH$_{3}$) being absent where a sulfur atom forms a chemical bond with the gold nano-cluster. 
SMMs can bond to a gold nano-cluster in various configurations, two of which we consider here: In Fig. \ref{geometry}(a) the magnetic easy axes of the two SMMs are aligned and are parallel to the z-axis whereas in Fig. \ref{geometry}(b) the two SMMs are misaligned so that their easy axes are not colinear.

Mn$_{12}$ SMMs have a degenerate ground state with spin S=10 and $S_{z}=\pm 10$ if the z-axis is chosen to be aligned with the easy axis.  
An applied magnetic field lifts this degeneracy and quantum tunneling of the magnetization (QTM) occurs between states on the opposite sides of the potential barrier (with $S_z > 0$ and $S_z<0$) leading to spin reversal.
The spin reversal mechanism depends on the temperature:
At zero temperature, QTM between the metastable ground state and the resonant energy level on the opposite side of potential barrier, shown by the blue arrow in Fig. \ref{geometry}(c), is followed by transitions to the ground state, giving rise to the magnetic reversal. Classical thermally activated processes (TAP) are responsible for the magnetization reversal at high temperatures as shown by red arrows in Fig. \ref{geometry}(c). 
These two processes, QTM and TAP, can act in concert resulting in thermally assisted tunneling in which QTM occurs between the energy levels near the top of potential barrier which are populated thermally, Fig. \ref{geometry}(d).

The hysteresis loop of  oriented Mn$_{12}$ crystal samples in the presence of a varying applied magnetic field shows steps at roughly constant field intervals \cite{hysteresis_steps}. The hysteresis staircase structures are due to QTM of SMMs at different applied magnetic fields.
Therefore, the QTM occurs at external magnetic fields $H_n=nH_1$ where $H_1=0.44T$ is a fixed parameter and $n$ is an integer.
Dynamic magnetic susceptibility  \cite{transvers_dynamic_susceptibility} and hysteresis measurements \cite{transvers_hysteresis} have been used to investigate the effect of a transverse magnetic field on the QTM and the relaxation time of SMMs.
These measurements reveal that $H_n$ depends only on the longitudinal magnetic field, $H_z=nH_1$. Therefore, by increasing the tilting angle between the applied magnetic field and the SMM's easy axis QTM occurs at a higher magnetic field. 
However, the increase of the transverse magnetic field increases the quantum tunneling probability and reduces the relaxation time.

In general, as in the case shown in Fig. \ref{geometry}(b), the two SMM's easy axes are not exactly aligned which leads to different tilting angles between the two SMM's easy axes and the external magnetic field. 
Then the two SMMs experience different longitudinal and transverse magnetic fields.
Experiencing different longitudinal magnetic fields
causes the magnetic moments of the two SMMs to flip at different external magnetic fields resulting in the possibility of a spin-valve effect in these molecular devices.
It is also worth mentioning that the spin reversal in SMMs is a quantum tunneling event. Therefore even for aligned configuration of the two SMMs, the spins of the two SMM's are expected to switch at different times.
   
At large negative external magnetic fields both SMMs will be in their ground states with positive spin quantum numbers; we will refer to this as the ``parallel configuration" (PC) whether the easy axes of the two SMMs are strictly colinear as in 
Fig. \ref{geometry}(a) or misaligned as in Fig. \ref{geometry}(b) . The system will remain in the PC until the external magnetic field switches direction; then at some point one of the SMMs will flip its spin direction and the system will go into the anti-parallel configuration (APC). Because of the energy barriers that hinder reversal of the magnetic moments of the SMMs, these parallel and antiparallel configurations can be maintained for some time after the magnetic field is set to zero at sufficiently low temperatures. Therefore, for simplicity, in what follows we will consider transport through the structures shown in Fig. \ref{geometry}(a) and \ref{geometry}(b) in the parallel and antiparallel spin configurations at zero externally applied magnetic field. 

Our SMM Hamiltonian is $H^{\mbox{\scriptsize{SMM}}}=H^{\mbox{\scriptsize{EH}}}+H^{\mbox
{\scriptsize{Spin}}}+H^{\mbox{\scriptsize{SO}}}$, as introduced in Ref. \onlinecite{PRB, PRBrapid}.
Here $H^{\mbox{\scriptsize{EH}}}$ is the extended H\"{u}ckel Hamiltonian.\cite
{extended_huckel} The spin Hamiltonian $H^{\mbox{\scriptsize{Spin}}}
$ gives rise to the magnetic polarization of the molecule. Spin orbit coupling is described by $H^
{\mbox{\scriptsize{SO}}}$. 
In extended H\"{u}ckel theory the basis is a small set of Slater-type atomic valence orbitals $|
\Psi_{i\alpha} \rangle$;  $|\Psi_{i\alpha} \rangle$ is the $i^\mathrm{th}$ orbital of the $\alpha^
\mathrm{th}$ atom. In this basis the extended H\"{u}ckel Hamiltonian elements are 
$H^{\mbox{\scriptsize{EH}}}_{i\alpha;{i}' {\alpha}'}= D_{i\alpha;
{i}' {\alpha}'} K (\varepsilon_{i\alpha}+\varepsilon_{{i}' {\alpha}'})/2$
where $D_{i\alpha;{i}' {\alpha}'} = \langle \Psi_{i\alpha} |\Psi_
{{i}' {\alpha}'} \rangle$ are orbital overlaps,
$\varepsilon_{i \alpha}$ is the experimentally
determined negative ionization energy of the $i^{th}$ valence orbital of the $\alpha^{th}$ atom, 
and $K$ is chosen empirically for consistency with experimental molecular electronic structure data.
In our calculations $K=1.75+\Delta_{i\alpha;{i}' {\alpha}'}^{2}-0.75\Delta_{i\alpha;{i}' {\alpha}'}^{4}$ where  $\Delta_{i\alpha;{i}' {\alpha}'}=({\varepsilon_{i\alpha}-\varepsilon_{{i}' {\alpha}'}})/({\varepsilon_{i\alpha}+\varepsilon_{{i}' {\alpha}'}})$ as was proposed in Ref. \onlinecite{huckel_off_diagonal}. 
The spin Hamiltonian matrix elements between
valence orbitals $i$ and $i'$ of atoms ${\alpha}$ and ${\alpha}'$ with spin $s$ and $s'$ are defined by
{\small
{\begin{equation*}\label{SpinHamiltonian} \begin{split}
 \langle i s \alpha |H^{\mbox {\scriptsize{spin}}}| {i}' s' {\alpha}' \rangle&={D_{i \alpha;{i}'{\alpha}'}}(\mathcal{A}_{i \alpha}+
\mathcal{A}_{i'{\alpha}'}) \langle s | \hat{n}\cdot {\bf{S}} | {s}' \rangle/{\hbar} \\
 \mathcal{A}_{i \alpha}&={\small \left\{\begin{matrix}
 \mathcal{A}_{\mbox{\scriptsize{inner}}} &\mbox {for inner Mn {\em d}-valence orbitals} \\ 
 \mathcal{A}_{\mbox{\scriptsize{outer}}} &\mbox {for outer Mn {\em d}-valence orbitals} \\ 
 0 &\mbox {otherwise} 
\end{matrix}\right.}
\end{split} \end {equation*}}}
Here $\hat{n}$ is a unit vector aligned with the magnetic moment of the SMM. $\bf{S}$ is 
the one-electron spin operator. $\mathcal{A}_{\mbox {\scriptsize{inner}}}$ and $\mathcal{A}_
{\mbox {\scriptsize{outer}}}$ are parameters chosen so that
the Hamiltonian gives rise to the correct spin for transition metal ions in the SMM ground state. 
 
\begin{figure}[t!]
\centering
 \setlength{\abovecaptionskip}{0pt plus 0pt minus 2pt}   
\includegraphics[width=1\linewidth]{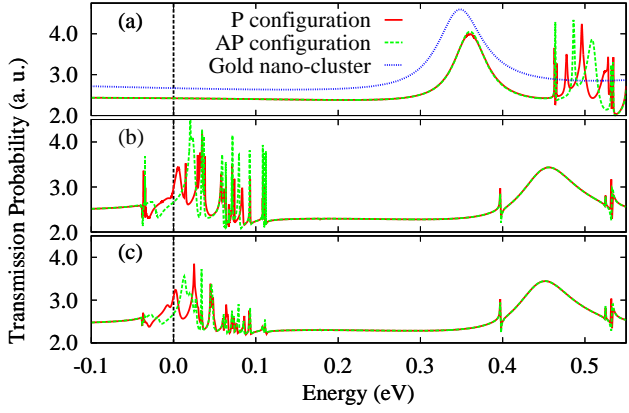}
\caption{(Color online) Calculated total electron transmission probabilities through the devices shown in Fig. \ref{geometry} (a) and \ref{geometry}(b).
Results for  parallel (P) and anti-parallel (AP) configurations of the magnetic moments of the two SMMs are red and green respectively. The case of aligned and misaligned easy axes as in Fig. \ref{geometry}(a) and \ref{geometry}(b) respectively shown in (a,b) and (c). (a) Gate potential V$_g= 0$ eV, (b-c) V$_g \approx  0.5$ eV is applied to both SMMs. The dotted vertical lines indicate the Fermi energy of gold. (a) The blue curve shows the transmission probability of the gold nano-cluster in absence of SMMs.} 
\label{transmission} 
\end{figure}

Spin-orbit coupling is responsible for the magnetic anisotropy of 
SMMs. We  
evaluate the matrix elements of the spin-orbit coupling Hamiltonian $H^{\mbox
{\scriptsize{SO}}}$ approximately from the standard expression\cite{Kittel}
$H^{\mbox{\scriptsize{SO}}}=\boldsymbol{\sigma}\cdot  \nabla{V(\bf{r}) \times \mathbf{p}  }~
{\hbar}/{(2mc)^2}$
where $\mathbf{p}$ is the momentum operator, $V(\bf{r})$ is the electron Coulomb potential 
energy, $\boldsymbol{\sigma} = (\sigma_x , \sigma_y, \sigma_z) $ and $\sigma_x , \sigma_y $ 
and $\sigma_z$ are the Pauli spin matrices. 
Evaluating the matrix elements of 
$H^{\mbox{\scriptsize{SO}}}$ between
valence orbitals $i$ and $i'$ of atoms ${\alpha}$ and ${\alpha}'$ with spin $s$ and $s'$ we  find
{\small
\begin{equation*} 
\begin{split}
\label{SOHamiltonianfinal}
\langle {i s \alpha} & | H^{\mbox{\scriptsize{SO}}}  | {{i}' s' {\alpha}'} \rangle \simeq 
 E^{\mbox{\scriptsize{SO}}}_{is{i}'s';\alpha}\delta_{\alpha \alpha'} \\
& +(1-\delta_{\alpha \alpha'})\sum_{j}(D_{i \alpha; j\alpha'}E^{\mbox{\scriptsize{SO}}}_{js i's'; 
\alpha'} 
 +[D_{i' \alpha'; j\alpha}E^{\mbox{\scriptsize{SO}}}_{js' is; \alpha}]^{\ast})
\end{split}
\end{equation*}}
The first term on the right is the intra-atomic contribution, the remaining terms are 
the inter-atomic contribution and
{\small
{\begin{equation*}   
\begin{split}
 \label{factor}
E^{\mbox{\scriptsize{SO}}}_{is{i}'s';\alpha}& =  \langle \alpha, l_i, d_i, s | \mathbf S \cdot \mathbf 
L_{\alpha} | \alpha, l_{i}, d_{i'}, s' \rangle  \\
& \times \langle R_{\alpha, l_i}  | \frac{1}{2m^2c^2} \frac{1}{|\mathbf{r}-\mathbf{r}_{\alpha}|} \frac
{dV(|\mathbf{r}-\mathbf{r}_{\alpha}|)}{d(|\mathbf{r}-\mathbf{r}_{\alpha}|)}  | R_{\alpha, l_{i}}  \rangle 
\delta_{l_i l_{i'}}
\end{split}
 \end{equation*}}}where the atomic orbital wave function $\Psi_{i \alpha}$   has been expressed 
as the product of a radial wave function $R_{\alpha, l_i}$ and directed atomic orbital $|\alpha, l_i, 
d_i, s \rangle$. Here $l_i$ is the angular momentum quantum number and $d_i$ may be $s, p_x, 
p_y, p_z, d_{xy},d_{xz},...$ depending on the value of $l_i$, while 
the radial integrals are the spin-orbit coupling constants.

This model describes the fundamental
properties of single Mn$_{12}$ SMMs quite well, yielding calculated
values of the SMM total spin, the magnetic moments of the inner
and outer Mn atoms, the magnetic anisotropy barrier and
the HOMO-LUMO gap that are consistent with experiment. \cite{PRB, PRBrapid}

Our transport calculations are based on Landauer theory and the Lippmann-Schwinger equation.
According to  the Landauer formula, $G=\frac{e^2}{h}T(E_{\mbox{\scriptsize{F}}})$, the conductance is proportional to the total transmission electron probability through the device at the Fermi energy  $E_{\mbox{\scriptsize{F}}}$ given by 
$ 
{\small T(E,V)= \sum_{ijs{s}'} \frac{v_{j{s}'}}{v_{is}}   \Big\vert t_{j{s}';is}} \Big\vert ^2
$. 
Here $E$ is the incident electron's energy, $t_{j{s}';is}$ is the transmission amplitude from $i^\mathrm{th}$  electronic channel of left lead with spin $s$  and velocity $v_{is}$ to $j^\mathrm{th}$ electronic channel of the right lead with spin ${s}'$ and velocity $v_{j{s}'}$.
We find the transmission amplitudes {\em{t}} by solving the Lippmann-Schwinger equation.

Fig. \ref{transmission}(a) shows the calculated total electron transmission probability through the device in Fig. \ref{geometry}(a)
as a function of the electron energy for the two SMMs with colinear easy axes and their magnetic moments in the parallel and anti-parallel configurations.  The PC and APC transmission probabilities are very similar for energies below $\sim 0.46~eV$. More importantly, the PC and APC transmission probabilities differ significantly at energies around $0.5~eV$, where the lowest unoccupied molecular orbitals (LUMOs) of the two SMMs and molecular orbitals (MOs) close in energy to the LUMOs are located.
[Note that the broad peak at energies near $0.38~eV$ is due to MOs located mainly on the gold nano-cluster as it is present also in the transmission probability of the gold nano-cluster in the absence of the SMMs, the blue curve in Fig. \ref{transmission}(a). For this reason 
the PC and APC transmission probabilities around this energy are very similar.]

\begin{figure}[t!]
\centering
 \setlength{\abovecaptionskip}{0pt plus 0pt minus 2pt}   
 \includegraphics[width=0.5\textwidth]{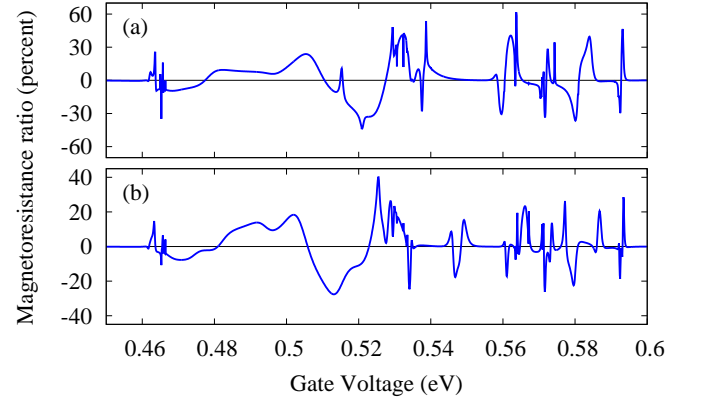}
\caption{ 
(Color online) Calculated magnetoresistance ratio (MR) as a function of applied gate voltage. The two SMMs attached  to the gold quantum dot have (a) colinear and (b) non-colinear easy axes as in Fig. \ref{geometry}(a) and \ref{geometry}(b) respectively.}
\label{transmission_probability_ratio}
\end{figure}

\begin{figure*}[t!]
\begin{center}
\begin{tabular}{c} 
 \includegraphics[width=1\textwidth]{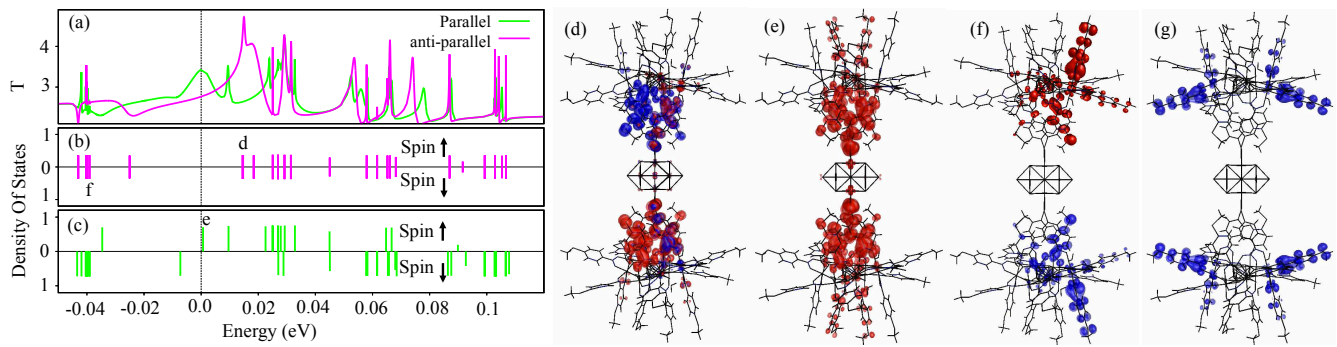}\\
\end{tabular}
 \end{center}
 \setlength{\abovecaptionskip}{0pt plus 0pt minus 2pt}   
\caption{
(Color online)Transmission probabilities (a), densities of states (DOS) (b,c) and representative molecular orbitals (d)-(g) of the device in Fig. \ref{geometry}(a). The DOS are projected onto carbon atoms of the system for (b) parallel and (c) antiparallel configurations of the two SMM's magnetic moments.  Gate voltage,$V_g \approx  0.5$ eV, is applied to the SMMs. Arrows in (b) and (c) indicate spin up/down. The axis of spin quantization is the z-axis shown in Fig. \ref{geometry}(a).
(d-g) Isosurface plots of molecular orbitals of this spin-valve system. Letters d, e, f, and g in (b) and (c) mark the corresponding MOs that are depicted below. Red (blue) represents the spin up (down) isosurface.
}
\label{DOS}
\end{figure*}

Fig. \ref{transmission}(b) presents the calculated PC and APC transmission probabilities of the same device as for  Fig. \ref{transmission}(a) but with a positive gate voltage applied to the SMMs; the gold nano-cluster is not gated. 
Here the gating has shifted the SMM LUMOs and MOs close in energy to the LUMOs to vicinity of Fermi energy of gold, indicated by the dotted vertical line in Fig. \ref{transmission}. 
Because the PC and APC transmission probabilities at the Fermi energy now differ significantly the system exhibits a substantial magnetoresistance ratio 
MR=(G$_{\mbox{\scriptsize{PC}}}$-G$_{\mbox{\scriptsize{APC}}}$)/G$_{\mbox{\scriptsize{APC}}}$. 
Since $G_{\mbox{\scriptsize{(A)PC}}}=\frac{e^2}{h}T_{\mbox{\scriptsize{(A)PC}}}(E_{\mbox{\scriptsize{F}}})$
, therefore MR=(T$_{\mbox{\scriptsize{PC}}}(E_{\mbox{\scriptsize{F}}})$-T$_{\mbox{\scriptsize{APC}}}(E_{\mbox{\scriptsize{F}}})$)/T$_{\mbox{\scriptsize{APC}}}(E_{\mbox{\scriptsize{F}}})$.

Note that although the transmission resonances around the Fermi energy in Fig. \ref{transmission}(b) resemble those around   $0.5~eV$ in Fig. \ref{transmission}(a) qualitatively, the gating has modified the {\em detailed} structure of the transmission peaks and dips, as should be expected for a mesoscopic quantum interference effect.
We find similar magnetoresistance ratios (MRs) for SMMs with non-colinear easy axes as is seen in Fig. \ref{transmission}(c) where we present our calculated PC and APC transmission probabilities for the device shown in Fig. \ref{geometry}(b) gated so as to bring the SMM LUMOs and nearby MOs close to the gold Fermi level. Similar MRs are expected regardless of the SMMs' misalignment angle.

Fig. \ref{transmission_probability_ratio} shows the magnetoresistance ratio 
of the colinear and non-colinear SMMs as a function of the gate voltage.
These results show that gating is able to transform such nanostructures into functioning spin valves controlled by the magnetic moments of the two SMMs that couple to the conducting path through the gold quantum dot via the SMM LUMOs and nearby MOs. 

Previous studies of Mn$_{12}$-Ph SMMs have shown the LUMO and MOs close in energy to the LUMO of these SMMs to have large amplitudes on the SMM's ligands and to be strongly spin polarized.\cite{PRB, PRBrapid}
In the molecular spintronic devices that we propose here, the strong presence of these spin polarized orbitals on the ligands and the strong electronic coupling between the ligands and the gold quantum dot allow delocalization of spin polarized MOs over entire system. Therefore, electronic transport through gold nano-cluster in the energy range occupied by the  LUMO (and MOs close in energy to the LUMO) is strongly affected by the relative orientation magnetic moments of the SMMs, as can be seen in  Fig. \ref{transmission}.

Fig. \ref{DOS}(a) shows the calculated total electron transmission probability through the device in Fig. \ref{geometry}(a)
as a function of the electron energy with a positive gate voltage applied to the SMMs in the parallel and anti-parallel configurations.
Fig. \ref{DOS}(b) and \ref{DOS}(c) show the densities of states (DOS) projected on the carbon atoms for the same systems as in Fig. \ref{DOS}(a) in parallel and anti-parallel configurations, respectively. Fig. \ref{DOS} shows the close correspondence between the DOS and resonant peaks in transmission probability plots. 
The spin resolved densities of states for the parallel and antiparallel configurations of the SMM magnetic moments are different. This results in the differing total electron transmission probability functions for the two configurations seen in  Fig. \ref{DOS}(a), and the consequent spin-valve functionality of the device.

In Fig. \ref{DOS}(a) the spin-valve effect is most pronounced in the energy range -0.02--0.03 eV where the transmission probabilities for PC and APC differ significantly over an extended range of energies. The broad spin valve features visible there are due to 
resonant and off-resonant transport through MOs which have significant weight on the ligands that connect the SMMs' magnetic cores to the gold nano-cluster. Fig. \ref{DOS}(d-e) show the wave function isosurfaces of such MOs in the parallel and anti-parallel configurations, respectively. These MOs are marked in the corresponding DOS plots by letters d and e.
As shown in  Fig. \ref{DOS}(d-e) these molecular orbitals  are located mainly on ligands that are connected directly to the gold nano-cluster. This direct connection favors a hybridization 
between SMMs and gold nano-cluster.
In cases where this hybridization occurs, resonant and off-resonant transport is expected.
On the other hand, these ligands bridge between the individual SMMs' magnetic cores and the gold nano-cluster, causing a strong interaction between the SMMs' magnetic cores which leads to the spin valve effect predicted by these calculations. By contrast, the transport resonances due to the MOs that are well separated (Fig. \ref{DOS}(f-g))  from gold nano-cluster are narrow and the calculated spin valve effect due to these resonances is confined to very narrow energy ranges.

In summary: Spin valve functionality, is a much sought property of molecular nano structures. We have proposed a molecular nano spin-valve prototype, based on transition metal Mn$_{12}$-Ph SMMs coupled to a gold quantum dot, and predict it to exhibit a large magnetoresistance spin-valve effect due to resonant transport. The transport resonances are mediated by electronic states that extend over the entire system due to hybridization between the gold cluster and spin-polarized SMM molecular orbitals with a strong presence on the SMM's ligands. We find the electronic transport through this system to depend strongly on the relative orientations of the magnetic moments of the SMMs grafted onto the gold nano-cluster. This work opens up a new avenue toward novel molecular spintronic devices based on transition metal SMMs.
 
This research was supported by CIFAR, NSERC, Westgrid and Compute Canada. 
We thank  B. Gates and B. L. Johnson for helpful comments and discussions.
 

\end{document}